\documentstyle[aas2pp4]{article}

\slugcomment{Not to appear in Nonlearned J., 45.}

\lefthead{Alonso et al.}
\righthead{Survey for ELGs: UCM List 3}

\begin{document}

\title{Survey for Emission-Line Galaxies: Universidad Complutense de Madrid 
List 3 \altaffilmark{1}}

\author{O. Alonso\altaffilmark{2}, 
C. E. Garc\'{\i}a-Dab\'{o}, J. Zamorano, J. Gallego, M. Rego}
\affil{Dpto. Astrof\'{\i}sica, Universidad Complutense de Madrid, 
E-28040 Madrid, Spain}
\authoremail{oal@eucmos.sim.ucm.es}



\altaffiltext{1}{Based on observations collected at the German-Spanish
Astronomical Center, Calar Alto, Spain, operated jointly by the Max-Planck
Institut f\"{u}r Astronomie (MPIA), Heidelberg, and the Spanish National
Commission for Astronomy.}

\altaffiltext{2}{present address: Dpto. Arquitectura de Computadores y
Autom\'{a}tica, Universidad Complutense de Madrid, E-28040 Madrid, Spain}

\begin{abstract}

A new low-dispersion objective-prism search for low-redshift ($z<0.045$) 
emission-line galaxies (ELG) has been carried out by the Universidad 
Complutense de Madrid with the Schmidt Telescope at the Calar-Alto 
Observatory.
This is a continuation of the UCM Survey, which was performed by 
visual selection of candidates in photographic plates via the 
presence of the H$\alpha+[NII]\lambda6584$ blend in emission. 
In this new list we have applied an automatic procedure, fully 
developed by us, for selecting and analyzing the ELG candidates on the 
digitized images obtained with the MAMA machine.
The analyzed region of the sky covers
189 square degrees in nine fields near $\alpha=14^h \ \& \  17^h, 
\delta=25\arcdeg$.
The final sample contains 113 candidates.
Special effort has been made to obtain a large amount of information 
directly from our uncalibrated plates by using several external calibrations. 
The parameters obtained for the ELG candidates allow for the study of the statistical properties for the sample.

\keywords{galaxies: general, surveys, starburst --- techniques: image 
processing --- Surveys --- }
\end{abstract}

\section{Introduction}
The Universidad Complutense de Madrid Survey (UCM Survey) has been carried
out during the last several years with the aim of looking for H$\alpha$
emission-line galaxies. It is described in detail by 
Zamorano et al. (1994, 1996), in UCM Lists 1 \& 2.

The UCM Survey was initiated with several objectives, the main
goals being: (1) to identify and study new young, low-metallicity galaxies;
(2) to carry out the classification and determination of the overall
properties and completeness of the sample of emission-line galaxies
(ELGs) selected; (3) to determine the spatial distribution and
luminosity function of the new galaxy population. We also wished to
compare our survey with others and to find out differences between the
sample obtained with various objective-prism techniques and to study
the overall relation between the far-infrared properties and the
optical behavior of the star-forming galaxies. Finally, we intended to
determine the evolutionary status and the different underlying stellar 
populations of the objects in order to detect any effect of evolution
in the starburst phenomena and to quantify the properties of the star
formation in the local universe.

The final product of Lists 1 and 2 was the Universidad Complutense de
Madrid (UCM) sample of star-forming galaxies in the local
Universe. This sample has been observed and analyzed in detail
(optical imaging: Vitores et al. (1996a, 1996b); spectroscopy:
Gallego et al. (1996, 1997); near infrared imaging: 
Alonso-Herrero et al. (1996), Gil de Paz et al. (1998)) 
and has provided the luminosity function for the star-forming galaxies
and the density of star formation rate in the local Universe (\cite{gal95}).

Lists 1 and 2 suffer from the subjective technique of looking for
candidates by visual inspection of the plates through a 10$\times$
binocular microscope. We intend to overcome this drawback by automatic
selection on the digitized plates. This method was shown to provide 
good performance and improvement in the selection of candidates by
applying it to several fields (\cite{oal95}; \cite{tesis}). In this work we
present the UCM List 3 obtained after using the automatic detection
process in nine contiguous fields covering 189 sq. deg. 

Section 2 describes the instrumental setup and observations.  In
section 3 the method used to select the candidates is
outlined. Section 4 presents several parameters measured for every 
galaxy of
the sample and explains the way they were obtained. In section 5 the
known selection effects are discussed. Finally, some 
statistical properties and
a comparison with other surveys are presented in section 6 and 7,
respectively.

\section{Observational Data}
This survey is based on photographic plates obtained with the 80/120
cm f/3 Schmidt Telescope of the Calar Alto German-Spanish Observatory
(Almer\'{\i}a, Spain) (\cite{bir84}) in June 1993. 
All the fields acquired for this research were obtained using 
the direct and
objective-prism configurations. In both modes, we use 8$\times$10
inch$^2$ plates, covering an useful field of 4.4$\times$5.5  deg$^2$
with a plate scale of 86$\arcsec$ mm$^{-1}$. They were hypersensitized
by baking them in an atmosphere of forming gas for 2 hr prior to
exposure. Full details of the plates are listed in table \ref{tbl-platedata}.

Direct plates were obtained exposing IIIaJ plates, combined with the
GG385 filter, for 1 hour. This configuration gives an instrumental
response similar to the Johnson B band. Prism plates were acquired in the red region of the spectrum. The 
use of IIIaF emulsion, with a sharp red
cutoff at 6850\AA, and a RG630 filter provides an useful spectral
range from $\sim$ 6400 to 6850\AA.
The objective-prism plates were obtained through a full aperture 
4$\arcdeg$ prism that yields a dispersion of 1980\AA \, mm$^{-1}$ at H$\alpha$ (see \S4.3). Dispersion of the prism spectra runs N-S axis. 
Our experience with the instrumental setup used indicates that exposures 
two hours long are a good balance between depth and plate background.
This instrumental configuration registers the
H$\alpha$+[NII]\altaffilmark{3} blend in emission for objects up to
z$\approx$0.045.

\altaffiltext{3}{ The combined H$\alpha$+[NII] emission will be referred to
in what follows as simply the H$\alpha$ emission.}

Quality standards for the analyzed plates are good seeing, no cloud
interference, no exposure interruption and telescope near the
meridian. Excellent guiding was achieved on this observing
run by the use of the new automatic guiding system installed on the
Schmidt Telescope.

The plates were scanned using the MAMA\altaffilmark{4} machine, a
high performance multichannel microdensitometer located at the
Observatoire de Paris (\cite{gui91}; \cite{mor92}). We used a pixel
size and sampling step of 10$\mu$m, the highest resolution achieved
by the machine. Each scanned plate returns a 23k$\times$18k pixel image
at 0.86$\arcsec$ pix$^{-1}$. 

\altaffiltext{4}{MAMA (Machine Automatique \`{a} Mesurer pour
l'Astronomie) is developed and operated by CNRS/INSU (Institut National des
Sciences de l'Univers) and located at the Observatoire de Paris}

Figure \ref{fig-sky} shows the fields covered by 
this discovery list. They are
placed in two regions located in the areas 
$13^h 45^m<\alpha<14^h 55^m$
; $21\arcdeg<\delta<27\arcdeg$ and $16^h 25^m<\alpha<17^h
50^m$ ; $21\arcdeg<\delta<27\arcdeg$. The total surveyed area is
189 $deg^2$.

\placetable{tbl-platedata}

\placefigure{fig-sky}

\section{Data reduction and selection of candidates}
In order to decrease the uncertainties intrinsic to visual scanning,
we have developed a new method for the automatic selection of
emission-line candidates in digitized objective-prism plates. The full
procedure is described in detail in Alonso et al. (1995) and 
Alonso (1996). Here we point out a brief discussion of the method.

The first task must be to find all the objects on the plates. To
locate and measure objects in the astronomical images we have
developed a program that performs this job by thresholding the image a
certain number of sigma over the local sky background. It returns a
catalogue with several parameters as positions, sizes, fluxes, etc. for
all the identified objects with a minimum selectable size. Our {\sc
BuhoFITS} software is able to handle large FITS files up to our
23k$\times$18k pixel images.  The large amount of spurious
identifications due to emulsion flaws, plate scratches or satellite
tracks are rejected by cross-correlating the catalogues obtained from
direct and prism plates for the same field.  We have used a plot of the
logarithm of the area versus density flux to discern between
stellar-like and extended objects (\cite{rei82}).

Subsequently the spectrum of 
all the objects are extracted by adding the
central five scans (i.e., those with higher signal), 
from the objective-prism image. These monodimensional
spectra are obtained in arbitrary and instrumental density units. No
calibration spots were recorded on the plates to obtain a photographic
density to intensity relation.  Nevertheless the emission feature is
clearly registered in these uncalibrated spectra and will be used in
all the subsequent analysis. Due to the spectrum extraction procedure
it can be noted that we are missing
those galaxies with emission out of the central region. This is one
of our selection effects, but it only affects very bright, extended
and well known galaxies.

The emission-line galaxy candidates are selected by analyzing the
prism spectra for all the objects. We use three different criteria in
order to perform this selection, as is described in detail in
Alonso et al. (1995).
In our low dispersion and uncalibrated plates, all the spectra show a 
similar appearance, mainly dominated by the instrumental response. Only 
objects showing the H$\alpha$ line in emission will present a clear 
discrepancy with respect to the continuum spectra of stars and galaxies 
without emission. The selection criteria are based in this fact, and
allow to compute a selection index for each object that
informs us about the reliability of the presence of an emission
feature. The final sample is obtained after visual inspection of those objects
with largest selection index. This step is needed because several
different configurations like overlapping stars, emulsion scratches in
the prism image, etc., can produce extracted spectra with an apparent
strong emission feature.

The UCM survey had a typical success of 70\% when looking for ELGs
(Lists 1 and 2). The remaining 30\% were confirmed to be objects
showing no emission. The automatic method has proved to improve these
results (\cite{oal95}; \cite{tesis}). In these studies we have compared 
the samples of ELG obtained after visual and automatic search in two 
fields. The new technique recovers nearly
80\% of the objects detected by visual scanning. All of the objects not recovered by the automatic method are extended and very bright galaxies or galaxies with
emission knots out of the nucleus.  Moreover, with the automatic
method we see an enhancement in the number of confirmed candidates
selected. 
As it is shown in Alonso et al. (1995) and Alonso (1996) an increase of 
24\% in the total number of objects
with confirmed H$\alpha$ emission has been reached with respect to the visual 
inspection.

\section{Physical Parameters}
We have put special attention into improving the extraction of
information from our uncalibrated photographic plates. Several
parameters have been obtained with intermediate accuracy for all the
candidates. Table \ref{tbl-ucmlist3} presents the full information for 
the objects of
the third list of our survey. Column (1) contains the name of the
galaxy according to the IAU rules. Equatorial coordinates for J2000.0
are given in columns (2) and (3). Column (4) presents an estimate of
the B magnitude obtained from our photographic plates (see \S4.4). In
columns (5) and (6) we present the size at the 25 mag 
arcsec$^{-2}$ isophote, in arcsec, and the position angle (see
\S4.2). Estimated redshifts are listed in column (7) (see \S4.3). 
Column (8) gives a preliminary morphological classification derived by visual
analysis of the Digitized Sky Survey\altaffilmark{5} (DSS) images, 
whereas column (9) classifies
each candidate into three categories according to the strength of the emission 
seen in the prism plates.
The meaning of the codes used in these columns are presented in the 
caption of the table.
Finally, column (10) shows previous designations of the
candidates. Finding charts from the DSS are
included in figure \ref{fig-chart}.

\altaffiltext{5}{The Digitized Sky Survey were produced at the Space
Telescope Science Institute under U.S. Government grant NAG
W-2166. The images of these surveys are based on photographic data
obtained using the Oschin Schmidt Telescope on Palomar Mountain and
the UK Schmidt Telescope. The plates were processed into the present
compressed digital form with the permission of these institutions. }

The parameters listed in the table are estimations, since 
spectroscopic and photometric studies have not been carried out.  Because 
our uncalibrated plates do not permit us to directly obtain absolute
parameters for the objects, we have performed several
external calibrations by relating our photographic parameters with
calibrated data for a small subsample of objects obtained from
different catalogues. However this approximation does not allow high
accuracy and some of the calibrations are not even 
possible. This occurs if
we try to measure the EW(H$\alpha$) in our spectra, for which we will
need a density to intensity calibration and, more important, a large
sample of emission-line galaxies with known EW(H$\alpha$) for each
plate in order to obtain a well sampled relation.

Next we briefly describe the procedure followed for obtaining each parameter.

\subsection{Astrometry}
We follow a standard procedure performed during the plate scan by the
MAMA machine in order to derive the constants for the plate
astrometry.  A third-order polynomial transformation is derived by
using astrometric reference stars from the PPM catalogue (\cite{roe91}).
The rms in the residuals of the astrometric reductions are about
0.3$\arcsec$, although large deviations can be achieved for very
extended objects with several components. Columns (2) and (3) of table
\ref{tbl-ucmlist3} catalogue the coordinates of the candidates referred to the equinox
J2000.0.

\subsection{Sizes and Position Angles}
During the object scanning process, we obtain the photographic major 
and minor axis and the position angle for each detected object by 
fitting an ellipse to pixels with a photographic density 3$\sigma$ 
over the local plate background. The sizes of a galaxy measured directly in
the plate cannot be straightforwardly converted to sizes in arc-seconds
via the plate scale because of the photographic plate behavior. An
ideal point of light is spread on the photographic emulsion over a
wide area, and this effect grows with the source brightness. Hence,
the size measured directly on the plate is a function of the real size
of the source, the magnitude, also the luminosity profile for an
extended source.  We have calibrated our photographic sizes with known
galaxies, using the Catalogue of Principal Galaxies (PGC) in its 
electronic version Extragalactic Card Index System (ECIS)
(\cite{pat89}). This catalogue has the advantage of being 
homogeneous, where
sizes are all referred to the isophote 25 mag arcsec$^{-2}$
(\cite{pat87}). In figure \ref{fig-size} we show the calibration derived 
for a single field. 
The other plates follow similar relations.

Figure also shows that the instrumental MAMA sizes are
systematically smaller than such obtained from the PGC. This fact is due to both the 
notably less depth of our photographic plates and the method used to obtain the sizes 
(thresholding the image 3$\sigma$ over the local background). Nevertheless, the 
calibrations derived allow us to estimate the size of our candidates within an 
uncertainty of 20$\arcsec$.
Higher precision could be achieved with more
homogeneous data. In Alonso (1996) an error of 2\arcsec \ was obtained
calibrating with  photometry for the UCM galaxies (\cite{al96b}).

The position angle is adopted as the orientation of the fitted
ellipse. There is no need for a calibration with external data, but a
checking was made using the same ECIS catalogue cited above. This
provided an estimation of the error, being 8\arcdeg \ for
1$\sigma$. For very elongated galaxies the error should be notably
smaller.

\placefigure{fig-size}

\subsection{Redshift}
The measurement of the redshift of the ELG candidates in our photographic
plates requires a wavelength calibration of the spectra. This
calibration involves obtaining the prism dispersion and requires the
knowledge of a reference point at a specific wavelength for each spectrum.

\placefigure{fig-po}

The prism disperses the light by means of the changing refractive
index with the following dependence on wavelength, in first order:

\begin{equation}
n(\lambda)= a + \frac{b}{\lambda^{2}}
\end{equation}

\noindent
where a,b are two constants. Thus, the position of a certain spectral
feature in the objective-prism image will follow the same dependency:

\begin{equation}
x(\lambda)= A + \frac{B}{\lambda^{2}}
\end{equation}

Here, the $A$ constant depends on the reference point we choose,
whereas the $B$ constant is only related to the characteristics of the
prism.

In order to obtain the dispersion curve for the 4\arcdeg \ objective
prism of the Calar-Alto Schmidt Telescope a test plate was taken.  The
configuration used was IIIaF emulsion without a filter. With a short
exposure the prism spectra of several bright stars were registered,
covering a spectral range from the blue atmospheric cut-off until
6850\AA \ due to the sharp red cut-off of the IIIa-F emulsion. 
The dispersion curve was obtained by
measuring the position of the strong absorption Balmer lines of A
stars. Fitting the data to the following expression (\cite{tesis}) (see 
figure \ref{fig-po})

\begin{equation}
\frac{d\lambda}{dx}=-\frac{\lambda^3}{2B} \\
\end{equation}

\noindent
yields a value of B=(-713 $\pm$ 2) 10$^8$ \AA$^2$ $\mu$m.
To measure the wavelength of the H$\alpha$ feature in the prism spectra
we also need the knowledge of a reference point. The emulsion red
cut-off at $\sim$ 6850\AA \ can not be used due to its high
dependency with brightness, color and size. We have used instead the
astrometric reduction to obtain a constant reference point for all the
spectra. This point is obtained by converting the position of the
object from the direct to the prism plate.
We do not know the
wavelength of this reference point, but it will be the same for all
the spectra, even for different plates, where we have applied the same
astrometric reduction, and it will not be affected by the effects
previously noted (size, brightness and color).
Only when the emission comes out of center of the galaxy the method 
clearly 
fails. Two identical galaxies showing H$\alpha$ emission, the first one 
in the nucleus 
and the second one in an external knot would show the emission line 
in different location of their prism spectra with respect to a 
same reference point, obtained from 
the plate to plate transformation. Therefore this fact would yield an 
erroneous value of the redshift for the second one.

In order to obtain the wavelength of this 
reference point, that is, the value of the constant $A$ we
represent the redshift ($z$) of known galaxies
with respect to the H$\alpha$ position in the prism spectrum, relative
to the reference point ($\Delta x$). This relation must follow the
expression:

\begin{equation}
z =  \frac{\sqrt[]{\frac{B}{\Delta x(\mu m)-A}}-\lambda_0}{\lambda_0}
\end{equation}

\noindent
where $\lambda_0$ is 6562.8\AA \ if we are working with the
H$\alpha$ line.  Therefore we can calculate the value of $A$ by
applying an algorithm of minimum squares to the data. The value of
such constant, obtained separately for the nine fields, yields the
same results inside 1$\sigma$ error bar, confirming the procedure.

Assuming an uncertainty of 1 pixel in the determination of the
center of the line, and taking into account the rms in the
calibration fit, the typical error in the computation of the redshift
results 0.004 for 1$\sigma$ (i.e. 1200 km s$^{-1}$).
In figure \ref{fig-calz} we present the data obtained applying the method just 
described for an small subsample of galaxies with known redshifts in
order to test our procedure. The galaxies showing large deviations 
are explained noting that the method may fail when the
galaxy is very extended and the emission feature comes from a
external knot. This is the case of UCM1436+2245 (IRAS 14360+2245),
with a redshift of 0.04 (\cite{fis95}). If we apply our method it
gives an estimate of z=0.017 due to the presence of a double component
emission probably originating in different knots. Column 7 of Table 
\ref{tbl-ucmlist3}
presents the redshift obtained for non-catalogued candidates. Data 
marked with an asterisk are obtained from literature.

\placefigure{fig-calz}

Obviously, these redshifts are based in the assumption that the emission 
feature registered in the prism spectra is the H$\alpha$ line. 
This conjecture is based 
in the fact that none of the confirmed emission-line galaxies 
from the previous UCM 
lists 1 \& 2 was revealed as distant galaxies with their emission 
lines redshifted to 
our spectral coverage. In addition, a galaxy with a M$^*=-21.4$ 
magnitude, located at 
z=0.3 will show the [OIII]5007 line in the H$\alpha$ region, but the 
apparent magnitude would
be notably fainter than the limit of our prism plates (m$_r \approx$ 18).

\subsection{Magnitudes}
The direct plates were obtained with an instrumental response near the
B band of the Johnson system, and therefore we are interested in
measuring the magnitudes of our ELG sample. Our plates are not
calibrated, but we can relate the photographic flux, the sum of the
density for all the pixels of the image, with the B magnitude for
objects recovered from databases. This relation works fine for stellar
objects ($\sigma \approx 0.1 \ mag$), but it is not straightforward to
apply to extended objects because the dependence with the size of the
source. Using both photographic density and size it is possible to
estimate the magnitudes, but with larger errors, up to 0.5 mag for
very extended galaxies.

Several studies have been published which have achieved better accuracy using
photographic plates, but in general all of them require calibration
spots (\cite{ESO}) or the calibration give good results only in
photometry for stellar-like objects (\cite{kroll}; \cite{berg91};
\cite{moh87}). In general, the photometry of extended sources has several
sources of error sometimes hard to correct for (see \cite{vac}).

B magnitudes listed in column (4) of table \ref{tbl-ucmlist3} are derived using the
method just described above, and should be used only as estimates for
statistical or observational purposes. Only values for compact galaxies
can be taken with greater confidence.

\section{Selection effects}

One of the major problems encountered when working with 
photographic plates is that the
characteristics of each of them can vary notably with respect to the
other ones, even taking special care in repeating all the
observational setup, being almost impossible to obtain a homogeneous
sample of galaxies. 
In order to understand the biases of our observational procedure, we 
have compared the samples of galaxies obtained in our survey with those
recovered by using different observational configurations. This job is 
presented in detail in Zamorano et al. (1994, 1996) for the UCM 
Lists 1 \& 2, and is also presented in section 7 of this paper for 
this new List 3.

\placefigure{fig-histoewf}

Nevertheless in this section we have used the spectroscopic data derived for
the two previous lists of the UCM survey (\cite{gal95}) with the aim of
investigating the observational biases of our instrumental
configuration. This analysis (\cite{tesis}) shows that neither 
the equivalent width (EW) of the
emission nor the flux of the H$\alpha$ line alone are the parameters that
controls the detection limit in prism surveys. Moss et al. (1988) introduced
an auxiliary parameter defined as EW$\times$Flux, and showed that it
defined accurately the observational limit of their work, an
objective-prism survey in the red region of the spectrum, at a
reciprocal dispersion of 400 \AA mm$^{-1}$.  We have also used this
parameter in our survey. In figure \ref{fig-histoewf} we 
show the histogram of
values EW(\AA)$\times$Flux(erg s$^{-1}$ cm$^{-2}$)
for the sample of UCM galaxies from Lists 1 \&
2. We can conclude that the UCM survey is able to identify
emission-line galaxies with a value of log(EW$\times$Flux) $>$ -13.
Only four objects have lower values of this parameter. Nevertheless 
three of them have redshifts larger than the limit imposed by the 
emulsion sensitivity, being impossible to register the H$\alpha$ 
emission. These objects must be considered as serendipitous 
identifications.

In figure \ref{fig-ewf} we show the distribution of the UCM objects of the 
Lists 1 \& 2 in a log(Flux) vs. log(EW) plot. The data have been
obtained from the work of Gallego et al. (1996).  The plot shows the behavior of
a sample obtained by prism plates. It also shows the lack of
galaxies in the upper-left and lower-right corners. These two regions
belong to very bright and faint galaxies respectively. Therefore the
diagram shows that, in addition to a limiting magnitude, there exists
a deficiency of bright galaxies due to saturation of the photographic
plates. We can not represent the location of the galaxies obtained
in this List 3 until performing spectroscopic observations.
Nevertheless, these effects are due not to the visual or automatic
procedure to select the candidates, but to the intrinsic
characteristics of the photographic emulsion. This is why we expect a
similar response for the galaxies identified in this list.

\placefigure{fig-ewf}

\section{Statistical properties\altaffilmark{6}} 

\altaffiltext{6}{For now on, we adopt a value of H$_{0}$=50 km s$^{-1}$
Mpc$^{-1}$ and q$_{0}$=0.5.}

\subsection{Apparent magnitudes}
The histogram of apparent magnitudes presented in figure \ref{fig-histom} informs us
about the depth  of the survey. Nearly all the galaxies lie between
16 and 18 in B magnitude, with a few reaching magnitude 20. The mean
apparent B magnitude for the sample is 16.8.  This histogram shows
a known bias of this technique, that is, it is unable to
detect bright galaxies because they appear saturated in the
photographic plates.  This occurs, for example, for several known
galaxies as NGC 5637 (m$_{B}$=14.7) or WAS 82 (m$_{B}$=15.5). 
We also have presented the histogram of Gunn-Thuan r magnitudes obtained
from the UCM Lists 1 \& 2.
Assuming a mean color B-r=0.65 for the UCM sample of galaxies, a typical value 
for Scd galaxies (Fukugita et al. 1995; Vitores et al. 1996b), the
comparison points out that 
UCM List 3 seems to be slightly deeper that the previous ones, although lack of precision in
our B magnitudes prevents us from extending these conclusions. We note that the 
large number of galaxies in the UCM List 1 \& 2 histogram is due to the greater
explored area compared with this List 3.

\placefigure{fig-histom}

\subsection{Absolute magnitudes}
Figure \ref{fig-histoabs} shows the
histogram in absolute magnitude. The mean
absolute magnitude is $M_B=-18.9$. This histogram is less symmetrical
than the distribution of $M_r$ derived from Lists 1 \& 2 (\cite{al96b}).
Our coverage of the low luminosity end of  the luminosity function is by 
far better than previous UCM lists.
It is worth noting than galaxies with $M_B$ as faint as -15
have been detected. Such percentage of low luminosity galaxies has
been also reached for the University of Michigan survey for 
emission-line galaxies (\cite{UM}).

\placefigure{fig-histoabs}

\subsection{Surface brightness}

We have computed the mean surface brightness using the apparent blue
magnitude and the size referred to the isophote 25 mag
arcsec$^{-2}$. 
The mean surface brightness of our sample is
distributed almost uniformly from 21.5 to 24.5 mag arcsec$^{-2}$ with
22.8 being the mean value (see figure \ref{fig-histosurf}).
Lists 1 \& 2 yield a mean surface brightness of 22 mag arcsec$^{-2}$ in the Gunn-Thuan r band (\cite{al96b}). The precision of our data prevents us to extend this study to filter or color differences. 

\placefigure{fig-histosurf}

In figure \ref{fig-surf} we plot B magnitude
vs. surface brightness for the
sample. The BCD type galaxies are mainly distinguished because of their
low luminosity and high surface brightness. The right side of the plot
is populated by such galaxies. 
Considering an upper limit of $M_B > -18$ (\cite{thuan81}) for a 
BCD galaxy, we have regarded those galaxies with a compact aspect and a more 
conservative $M_B > -16.5$ as good candidates to be BCD galaxies
Attending to this criteria
the most probable candidates
for BCD's are: UCM1345+2417, UCM1413+2446, UCM1449+2559, 
UCM1735+2617, UCM1742+2343 and UCM1742+2634. Other type 
of galaxies found in this
region are irregular galaxies with strange morphologies such as
comet--like shapes, double--component, etc. Such galaxies are
UCM1640+2238, UCM1721+2326, UCM1723+2556 and so they were 
not included in the BCD sample.

\placefigure{fig-surf}

\subsection{Spatial distribution}
In figure \ref{fig-pie} we show the spatial 
distribution of our sample for both the
14$^h$ and the 16$^h$ regions. The dots represent galaxies from the
CfA survey (\cite{huc95}) which have been plotted to show for the
normal galaxy distribution. It can be seen that several clustering
structures appear in the CfA data. A well-defined cluster is at 5000
km/s and three more at 9000 km/s, 9500 km/s and 11000 km/s 
(beyond the UCM instrumental limit in redshift) in the 14$^h$ field. 
In the
$17^h$ sector only one cluster is observed at 10000 km/s. At such redshift, we detect no galaxies.

\placefigure{fig-pie}

We have presented the location of the UCM galaxies using open circles.
They nearly follow the distribution of the CfA survey, 
althogh less clustering seems to appear.
Preliminary analysis points out the idea that preferred galaxy location 
could be related with physical size.

\section{Comparison with other surveys}
The UCM Survey has found 113 candidates in the 9 fields (189
deg$^{2}$) of this List 3. The overall density is around 0.59
candidates per square degree, slightly greater that the value
derived from UCM Lists 1 \&
2. It is worth noting that the actual density of galaxies with
H$\alpha$ emission (excluding candidates with no emission) is $\sim$
0.4 for the first lists (\cite{gal96}). One of the properties of the
automatic method is a better success rate of detection
(\cite{tesis}) (which will be determined for this survey with follow-up
spectroscopy).  Therefore this value is consistent with an improvement
in the number of galaxies detected. Although we have lost 
some bright and extended
galaxies, at fainter magnitudes the automatic procedure is more 
sensitive, providing an increase in the number of galaxies per
square degree.

In Lists 1 \& 2 some comparisons were made between the samples
detected with different surveys and observational techniques, and that
obtained by us. The conclusions we derived do not change when the new
UCM fields surveyed are added to the comparison. Following
Kinman (1984) the CGCG galaxies have been used as the reference sample
of galaxies in the field. Only 26 galaxies out of 242 galaxies with
known redshift z$\leq$0.04 have been found with emission. The ratio is
$\sim$11\%, similar to the 13\% found in the previous lists.

The UCM List 3 fields do not overlap completely with the KUG survey
(Kiso ultraviolet-excess galaxies, Takase et al. (1993)). If we restrict our
analysis to the common region ($13^{h}45^{m}-14^{h}30^{m}$), the KUG
survey has found 94 objects and UCM 26 candidates; only 7 have been
selected by both surveys. Thus only 27\% of UCM galaxies are also
KUG objects. Since the KUG survey selects objects by their colors, it
is not redshift limited. We expect that a fraction of the KUGs have a
redshift that prevents detection by us. For comparison Comte et al. (1994)
found 25\% of KUGs with $z>0.04$. The number density of KUGs is 1.8
objects per square degree for the total survey and outnumber the
surface density of UCM by a factor of 3.

There are 5 galaxies of the Wasilewski (1983) catalogue of emission-line
galaxies that have been also detected by our survey. Although the
detection technique is similar (objective-prism), he used IIIaJ
emulsion to register H$\beta$ and [OIII] lines. The one galaxy
undetected by us is WAS82 (m$_B$=15.5) due to a saturation problem
(see \S6).

\section{Summary}
We have presented the UCM survey List 3. This new research is the
scientific continuation of the previous two lists, but in this last
list we have applied an automatic procedure developed and tested by us
in a previous work in order to improve the results obtained up to now
with the visual search of candidates. This List 3 covers 189 sq. deg.
in nine fields, and yields a total number of 113 candidate
H$\alpha$ emission-line galaxies, giving an overall density of 0.59
candidates per square degree.

The nine fields were acquired in direct and objective-prism modes, and
the plates were scanned using the MAMA machine. This procedure has
permitted us to recover a great amount of information directly from
the plates, allowing to perform statistical analysis of the sample
before carrying out photometric or spectroscopic observations. We
have derived high precision coordinates, magnitudes in the
blue band, sizes and redshifts for the whole sample of candidates, all
with moderate and known errors. Because we work with
uncalibrated plates, all these parameters are obtained by comparing our
photographic data with several parameters for objects recovered from
various databases.  These external calibrations allow for the
determination of the parameters noted above.

The determination of the dispersion curve of the prism used in the
survey has permitted us to estimate the redshift for the whole sample of
candidates by measuring the position of the H$\alpha$ line in the
prism spectrum related to the location of a constant reference point
obtained using the astrometric calibration. 
The $1-\sigma$ error in the computation of the redshift by this method 
results 0.004 (1200 km s$^{-1}$).
The comparison with known objects shows the precision of the method 
developed.

The photometric data obtained for the sample (apparent and absolute
magnitudes and sizes) follow the behavior derived for the two previous
lists using specific photometric observations. This result suggests that the automatic procedure selects a sample 
of galaxies comparable with those obtained in the previous lists.

The success of the new automatic procedure adopted in this new third
list is both the establishment of objective criteria for the selection 
of candidates and the extraction of several quantitative parameters,
in comparison with the lack of such information from previous visual-based  lists.

\acknowledgments We would like to gratefully acknowledge the
inestimable observation support received from the Calar Alto
Observatory staff, specially from Kurt Birkle. We also express our
thanks to Jean Guibert and the MAMA staff for their friendly
assistance and warm hospitality. We would like to thank to Armando Gil
de Paz, Javier Cenarro and Nicol\'as Cardiel for their helpful comments
and valuable suggestions. We would also like to thanks the anonymous 
referee for his useful suggestions and comments that improved this paper.
This work has made use of the NASA/IPAC
Extragalactic Database (NED), which is operated by the Jet Propulsion
Laboratory, Caltech, under contract with the National Aeronautics and
Space Administration. This research was supported in part by the
Spanish Programa Sectorial de Promoci\'{o}n General del Conocimiento
under grants PB93--456, PB96--0065, PB96--0610.

\placetable{tbl-ucmlist3}

\placefigure{fig-chart}

\clearpage

\begin{figure}
\caption{Schematic map of the sky showing the field
covered by the UCM List 3. Field centers are provided in table 
\ref{tbl-platedata}.
\label{fig-sky}}
\end{figure}

\begin{figure}
\caption{Relationship between the sizes for a small sample of galaxies
taken from ECIS catalogue and those 
measured in our direct plates. Major and
minor axis values are represented by squares and crosses
respectively. The straight line represents the best fit. See \S4.2 for
details.\label {fig-size}}
\end{figure}

\begin{figure}
\caption{Dispersion curve for the 4$\arcdeg$ objective prism of 
the Schmidt Telescope at the Calar Alto Observatory. 
The bottom plot 
shows the position of some spectral features measured in A-type stars 
relative to the H$\beta$ line vs. wavelength. The curve displays the 
best fit assuming a refraction index law n($\lambda$)=a+b/$\lambda^2$. 
The top diagram shows the dispersion dependence with wavelength. \label{fig-po}}
\end{figure}

\begin{figure}
\caption{Relationship between the redshift obtained from the prism spectra,
applying the method described in \S4.3, and those recovered from the literature for a small subsample of ELGs.\label{fig-calz}}
\end{figure}

\begin{figure}
\caption{Histogram of values EW$\times$Flux (of the H$\alpha$ emission line) for the sample of galaxies from UCM Lists 1 \& 2.\label 
{fig-histoewf}}
\end{figure}

\begin{figure}
\caption{H$\alpha$ Flux vs. EW for the sample of galaxies from 
UCM Lists 1 \& 2. Solid line connect the points with the parameter 
EW$\times$Flux constant. Dashed lines show the location of galaxies 
with continuum flux (in erg s$^{-1}$ cm$^{-2}$ \AA$^{-1}$) constant.
\label{fig-ewf}} 
\end{figure}

\begin{figure}
\caption{Histograms of apparent magnitudes from UCM Lists 1 \& 2 (dashed
line; Gunn-Thuan r magnitudes) and UCM List 3 (solid line; B magnitudes).
\label{fig-histom}}
\end{figure}

\begin{figure}
\caption{Histogram of absolute B magnitudes obtained for UCM List 3.
\label {fig-histoabs}}
\end{figure}

\begin{figure}
\caption{Histogram of mean B surface brightness inside the 25
mag/arcsec$^2$ isophote.\label {fig-histosurf}}
\end{figure}

\begin{figure}
\caption{Surface brightness vs. absolute magnitude in the B band. 
Dashed lines represent galaxies with constant diameter size.
Blue Compact Galaxy candidates are marked with a star.\label {fig-surf}}
\end{figure}

\begin{figure}

\caption{Redshift-position diagrams for candidates of UCM List 3.
The slices cover 6\arcdeg \ centered at 24.5\arcdeg \ in declination.
Dots represent galaxies from the CfA survey while open circles show the UCM galaxies.\label {fig-pie}}
\end{figure}

\begin{figure}
\caption{Finding charts for UCM galaxies of List 3. Each field covers
a square 5\arcmin \ wide. Object is centered in the frame.\label{fig-chart}} 
\end{figure}


\clearpage

\begin{deluxetable}{ccccccr}
\scriptsize
\tablecolumns{7}
\tablewidth{0pc}
\tablecaption{Plate Data \label{tbl-platedata}}
\tablehead{
\colhead{Plate Ident.} & \multicolumn{2}{c}{Plate Center (J2000)} & \colhead{Common} &
\colhead{Seeing (\arcsec)} & \multicolumn{2}{c}{Number of} \\
\colhead{OP/Direct} & \colhead{RA} & \colhead{DEC} & \colhead{Field} & \colhead{OP/Direct} & \colhead{Objects} & \colhead{ELGs} }
\startdata
A516/A503 & 14 46 56 & +23 59 27 & 4.15\arcdeg $\times$ 5.14\arcdeg & 1.0/1.5 & 14213  & 13 \nl
A513/A510 & 14 29 15 & +23 53 45 & 4.24\arcdeg $\times$ 5.20\arcdeg & 
2.5/1 .0& 14932  & 14 \nl
A495/A507 & 14 13 12 & +23 49 25 & 3.92\arcdeg $\times$ 5.04\arcdeg &
 1.5/1 .0& 16999  &  9 \nl
A497/A506 & 13 56 38 & +23 57 39 & 4.06\arcdeg $\times$ 5.13\arcdeg & 2.0/1.0   & 17321  & 25 \nl
A496/A505 & 17 40 52 & +24 23 34 & 4.15\arcdeg $\times$ 4.92\arcdeg & 1.5/2 .0& 85601  & 19 \nl
A498/A509 & 17 23 17 & +24 33 57 & 4.15\arcdeg $\times$ 5.05\arcdeg & 2.0/1.0   & 75618  & 13 \nl
A500/A504 & 17 05 55 & +24 30 09 & 4.07\arcdeg $\times$ 5.15\arcdeg & 1.0/1.0   & 47559  & 10 \nl
A514/A508 & 16 50 01 & +24 04 37 & 4.13\arcdeg $\times$ 5.17\arcdeg & 2.5/1 .0& 27558  &  5 \nl
A517/A512 & 16 32 24 & +24 04 30 & 4.07\arcdeg $\times$ 5.13\arcdeg & 1.0/1.0   & 29583  &  5 \nl
\enddata
\normalsize
\end{deluxetable}

\newpage

\begin{deluxetable}{ccclcrlccl}
\scriptsize
\tablewidth{0pt}
\tablecaption{UCM Survey List 3 \label{tbl-ucmlist3}}
\tablehead{
\colhead{UCM} &  \colhead{RA}  &   \colhead{DEC} &  \colhead{ m$_B$} &   \colhead{Size} &  \colhead{PA}  &   \colhead{Redshift} &   \colhead {Morph} &
\colhead {OP emission} & \colhead{Other names} \nl
\colhead{(1)} &   \colhead{(2)} &   \colhead{(3)} &   \colhead{(4)}   &   \colhead{(5)}  &   \colhead{(6)} &   \colhead{(7)}      &   \colhead{(8)}
&   \colhead{(9)} &   \colhead{(10)}}
\startdata
1345+2417 &    13 48 06.3 &   24 02 21 &   17.0 &   18$\times$14 &   143 &   0.004 & C & w&   \nl
1345+2457 &    13 47 36.5 &   24 42 13 &   17.0 &   18$\times$17 &   0 &   0.024 & I & m&   \nl
1346+2420 &    13 49 18.2 &   24 05 45 &   16.0 &   37$\times$18 &   111 &   0.021 & I & m&   \nl
1347+2527 &    13 49 37.6 &   25 13 04 &   18.0 &   8$\times$7 &   72 &   0.023 & * & m&   \nl
1348+2147 &    13 51 17.8 &   21 32 38 &   15.2$*$ &   35$\times$30 &    99 &   0.025 & S$_f$ & w&   KUG 1348+217 \nl
1348+2510 &    13 50 20.2 &   24 56 09 &   17.0 &   24$\times$12 &   2 &   0.031 & S & s&   IRAS F13480+2511 \nl
1349+2151 &    13 51 24.8 &   21 36 19 &   17.0 &   14$\times$11 &   58 &   0.038 & * & s&   \nl
1349+2152 &    13 51 38.0 &   21 37 39 &   16.5 &   36$\times$13 &   161 &   0.031 & S$_e$ & m&   IRAS F13492+2152 \nl
1350+2207 &    13 53 20.3 &   21 53 09 &   16.5 &   21$\times$13 &   172 &  0.033 & I & m&   \nl
1350+2456 &    13 52 26.5 &   24 41 29 &   19.0 &   6$\times$4 &   105 &   0.029 & * & m&   \nl
1350+2529 &    13 52 24.9 &   25 14 47 &   17.0 &   17$\times$12 &   53 &   0.032 & O & m&   \nl
1351+2201 &    13 53 25.7 &   21 46 17 &   17.0 &   14$\times$11 &   108 &   0.036 & O & s&   \nl
1351+2521 &    13 53 38.7 &   25 06 41 &   17.0 &   20$\times$11 &   109 &   0.027 & S$_e$ & w&   \nl
1352+2202 &    13 54 58.8 &   21 48 17 &   16.0 &   24$\times$19 &   29 &   0.036 & S$_f$ & m&   \nl
1352+2256 &    13 54 39.2 &   22 41 40 &   17.0 &   22$\times$15 &   51 &   0.023 & S$_f$& m&   \nl
1353+2507 &    13 55 37.3 &   24 52 37 &   17.0 &   16$\times$12 &   82 &   0.030 & O & w&   \nl
1353+2517 &    13 55 34.4 &   25 02 59 &   16.09$*$ &   35$\times$21 &   21 &   0.0295$*$ & S$_f$ & w&   CGCG 132-048 \nl
1353+2531 &    13 55 36.1 &   25 16 27 &   17.0 &   17$\times$9 &    97 &   0.023 & S$_e$ & m&   \nl
1353+2647 &    13 55 38.0 &   26 32 51 &   17.02$*$ &   16$\times$10 &   112 &   0.009 & S$_f$& w&   NPM1G +26.0340 \nl
1355+2440 &    13 57 37.8 &   24 26 04 &   17.0 &   15$\times$10 &   69 &   0.032 & C & m&   \nl
1356+2157 &    13 58 36.2 &   21 43 16 &   16.0 &   46$\times$16 &   18 &   0.032 & S$_e$& s&   \nl
1356+2310 &    13 58 24.7 &   22 55 39 &   15.5 &   29$\times$21 &   127 &   0.017 & S$_f$ & m&   \nl
1357+2614 &    14 00 12.0 &   26 00 21 &   18.5 &   12$\times$8 &   108 &   0.020 & I & m&   \nl
1400+2304 &    14 03 05.9 &   22 50 26 &   17.41$*$ &   28$\times$13 &   57 &   0.019 & S & w&   NPM1G +23.0349 \nl
1401+2602 &    14 04 01.8 &   25 47 44 &   14.9$*$ &   18$\times$15 &   115 &   0.0330$*$ & O & s&   WAS 89 \nl
1402+2152 &    14 04 53.0 &   21 38 09 &   14.9$*$ &   38$\times$29 &    95 &   0.0165$*$ & S & w&   MRK 0667 \nl
1408+2543 &    14 10 57.2 &   25 29 48 &   14.43$*$ &   71$\times$64 &    92 &   0.0316$*$ & S$_f$& w&   IC 4381 \nl
1408+2547 &    14 10 54.2 &   25 33 14 &   15.9$*$ &   20$\times$16 &   84 &   0.0312$*$ & S$_f$ & m&   WAS 90 \nl
1408+2623 &    14 10 28.5 &   26 09 05 &   16.0 &   18$\times$17 &   48 &   0.029 & C & m&   \nl
1413+2317 &    14 15 22.3 &   23 03 50 &   17.0 &   22$\times$13 &   106 &   0.023 & S & m&   \nl
1413+2446 &    14 15 27.3 &   24 32 26 &   20.0 &   7$\times$5 &   118 &   0.024 & * & m&   \nl
1416+2202 &    14 18 42.9 &   21 49 09 &   15.0 &   78$\times$38 &   164 &   0.031 & I$_{IP}$ & s&   UGC 09164 \nl
1416+2300 &    14 19 07.5 &   22 46 19 &   17.15$*$ &   19$\times$18 &   175 &   0.019 & S & m&   NPM1G +23.0356 \nl
\tableline
\tablebreak
1416+2543 &    14 18 25.4 &   25 30 04 &   15.7$*$ &   62$\times$18 &   28 &   0.0150$*$  & S$_e$  & w&   KUG 1416+257 \nl
1418+2209 &    14 20 46.6 &   21 56 14 &   14.3$*$ &   16$\times$32 &   154 &   0.0156$*$ & S$_e$ & m&   UGC 09182 \nl
1419+2420 &    14 21 52.9 &   24 06 27 &   15.6$*$ &   30$\times$20 &   72 &   0.020 & S & m&   KUG 1419+243 \nl
1422+2321 &    14 25 00.0 &   23 07 32 &   16.0 &   23$\times$22 &   106 &   0.017 & S$_f$ & s&   KUG 1422+233 \nl
1422+2450 &    14 24 22.9 &   24 36 52 &   14.11$*$ &   93$\times$45 &   134 &   0.0171$*$ & S$_B$ & m&   NGC 5610 \nl
1424+2515 &    14 26 26.8 &   25 01 47 &   17.0 &   14$\times$12 &    94 &   0.022 & O & w&   \nl
1424+2537 &    14 26 19.8 &   25 24 03 &   15.5$*$ &   31$\times$23 &   82 &   0.019 & O & w&   KUG 1424+256 \nl
1424+2541 &    14 26 15.5 &   25 27 59 &   16.5 &   14$\times$13 &    94 &   0.017 & * & w&   \nl
1425+2146 &    14 27 34.1 &   21 33 25 &   17.0 &   21$\times$11 &   24 &   0.030  & S & w&   \nl
1426+2322 &    14 29 10.9 &   23 08 55 &   18.0 &   10$\times$10 &   108 &   0.023 & C & w&   \nl
1427+2314 &    14 30 11.0 &   23 01 36 &   15.3$*$ &   39$\times$32 &   120 &   0.0173$*$ & S & w&   MRK 0683 \nl
1429+2145 &    14 31 20.9 &   21 32 10 &   16.5 &   25$\times$13 &   118 &   0.018 & S & w&   \nl
1431+2441 &    14 33 20.3 &   24 28 04 &   17.0 &   27$\times$12 &   33 &   0.034 & S & w&   \nl
1432+2550 &    14 35 06.5 &   25 37 48 &   16.5 &   16$\times$12 &    93 &   0.015 & S & w&   \nl
1435+2249 &    14 38 10.4 &   22 36 27 &   16.5 &   14$\times$14 &   17 &   0.025 & S & w&   \nl
1436+2245 &    14 38 21.1 &   22 32 12 &   16.5 &   27$\times$18 &   11 &   0.017 & I & w&   IRAS 14360+2245 \nl
1437+2148 &    14 39 58.5 &   21 35 59 &   16.5 &   36$\times$15 &   35 &   0.031 & S & m&   LSBC F580-06 \nl
1438+2209 &    14 41 15.8 &   21 56 44 &   16.5 &   20$\times$19 &   169 &   0.025 & S$_f$ & m&   NPM1G +22.0467 \nl
1438+2239 &    14 40 54.9 &   22 27 08 &   17.0 &   22$\times$12 &   144 &   0.014 & S & w&   IRAS F14386+2239 \nl
1438+2307 &    14 41 15.2 &   22 54 32 &   17.5 &   13$\times$12 &   164 &   0.0339$*$ & S & m&   IRAS 14389+2307 \nl
1440+2521 &    14 43 02.7 &   25 09 08 &   16.16$*$ &   28$\times$14 &   47 &   0.0319$*$ & S & s&   UGC 09489 \nl
1442+2248 &    14 44 35.5 &   22 35 38 &   17.5 &   14$\times$13 &   120 &   0.025 & S & m&   \nl
1446+2312 &    14 48 45.2 &   22 59 34 &   15.5 &   45$\times$22 &   111 &   0.008 & S$_{IP}$ & w&   IRAS F14465+2311 \nl
1447+2535 &    14 49 35.8 &   25 22 52 &   14.34$*$ &   46$\times$45 &    99 &   0.0339$*$ & S$_f$ & m&   UGC 09544 \nl
1448+2248 &    14 50 38.5 &   22 36 31 &   17.5 &   12$\times$12 &   27 &   0.034 & C & w&   \nl
1448+2256 &    14 50 37.8 &   22 44 06 &   15.7$*$ &   25$\times$22 &   168 &   0.0215$*$ & S & s&   MRK 1388 \nl
1449+2559 &    14 51 33.2 &   25 46 58 &   19.0 &   7$\times$6 &   4 &   0.021 & C & m&   \nl
1450+2342 &    14 52 24.1 &   23 30 41 &   17.5 &   14$\times$11 &   51 &   0.034 & S & m&   \nl
1624+2359 &    16 26 23.7 &   23 52 41 &   17.0 &   20$\times$14 &   22 &   0.040 & S & m&   IRAS 16242+2359 \nl
1627+2433 &    16 29 52.8 &   24 26 39 &   15.5$*$ &   35$\times$34 &   142 &   0.0375$*$ & I & m&   VV 807 \nl
1628+2453 &    16 30 55.8 &   24 46 49 &   17.0 &   20$\times$16 &    98 &   0.026 & S & w&   \nl
1636+2632 &    16 38 02.6 &   26 27 04 &   15.93$*$ &   25$\times$22 &   29 &   0.009 & S & m&   NPM1G +26.0432 \nl
1637+2417 &    16 39 26.1 &   24 11 59 &   17.48$*$ &   18$\times$15 &   140 &   0.015 & C & w&   NPM1G +24.0414 \nl
1640+2238 &    16 42 38.5 &   22 33 10 &   17.5 &   32$\times$11 &   37 &   0.011 & I & m&   \nl
1640+2510 &    16 42 23.8 &   25 05 07 &   14.8$*$ &   89$\times$28 &   161 &   0.0227$*$ & S & m&   UGC 10514 \nl
1643+2213 &    16 45 15.0 &   22 08 22 &   15.7$*$ &   41$\times$23 &   12 &   0.0316$*$ & S & w&   CGCG 138-069 \nl
1647+2259 &    16 49 23.9 &   22 54 16 &   17.0 &   15$\times$11 &   179 &   0.025 & * & m&   \nl
1650+2551 &    16 52 31.8 &   25 46 25 &   15.94$*$ &   28$\times$22 &   139 &   0.0348$*$ & S & s&   IRAS 16504+2551 \nl
1655+2532 &    16 57 23.4 &   25 27 57 &   16.61$*$ &   20$\times$16 &   109 &   0.040 & S & m&   NPM1G +25.0438 \nl
1656+2413 &    16 58 33.1 &   24 08 51 &   18.0 &   17$\times$11 &   166 &   0.019 & I & m&   \nl
1656+2450 &    16 58 47.0 &   24 46 24 &   17.5 &   38$\times$12 &   42 &   0.025 & S & w&   \nl
1701+2535 &    17 03 05.1 &   25 31 48 &   17.0 &   16$\times$13 &   163 &   0.039 & * & s&   \nl
1701+2642 &    17 03 48.2 &   26 38 37 &   17.0 &   12$\times$11 &   152 &   0.027 & * & s&   \nl
1702+2314 &    17 04 59.9 &   23 10 10 &   16.0 &   34$\times$25 &   72 &   0.0304$*$ & S$_f$ & w&   CGCG 139-033 \nl
1706+2300 &    17 08 52.3 &   22 57 10 &   17.0 &   18$\times$10 &   46 &   0.022 & * & s&   \nl
1710+2316 &    17 12 45.9 &   23 13 28 &   14.15$*$ &   53$\times$38 &   33 &   0.034 & S$_B$ & s&   NGC 6315 \nl
1711+2427 &    17 13 12.3 &   24 23 42 &   19.0 &   14$\times$6 &   110 &   0.035 & I & w&   \nl
1712+2305 &    17 14 25.5 &   23 01 39 &   17.0 &   20$\times$12 &   153 &   0.030 & S & w&   \nl
1712+2306 &    17 14 30.0 &   23 03 38 &   15.1$*$ &   34$\times$27 &   48 &   0.0295$*$ & S & w&   ARK 520 \nl
1714+2442 &    17 16 54.3 &   24 38 54 &   18.0 &   14$\times$13 &   32 &   0.021 & I & m&   \nl
1714+2541 &    17 16 42.6 &   25 38 03 &   15.7$*$ &   39$\times$22 &   116 &   0.023 & S & w&   CGCG 140-004 \nl
1717+2428 &    17 19 34.6 &   24 25 31 &   18.0 &   12$\times$12 &   15 &   0.028 & O & s&   \nl
1717+2458 &    17 19 56.6 &   24 55 57 &   17.0 &   29$\times$18 &   126 &   0.019 & S$_B$ & s&   \nl
1721+2326 &    17 23 29.0 &   23 23 36 &   17.5 &   25$\times$22 &   23 &   0.005 & I$_{IP}$ & s&   \nl
1722+2500 &    17 24 45.4 &   24 58 17 &   14.2$*$ &   68$\times$37 &   43 &   0.0276$*$ & S & w&   UGC 10837 \nl
1722+2656 &    17 24 50.7 &   26 53 32 &   17.0 &   29$\times$26 &   144 &   0.031 & S$_f$ & w&   \nl
1723+2556 &    17 25 48.6 &   25 53 33 &   18.0 &   17$\times$11 &   135 &   0.014 & C$_{IP}$ & w&   \nl
1725+2653 &    17 27 47.0 &   26 51 16 &   15.74$*$ &   31$\times$26 &   161 &   0.0296$*$ & S & w&   VV 389 \nl
1726+2339 &    17 28 18.8 &   23 37 27 &   15.5 &   30$\times$23 &   62 &   0.030 & S & s&   CGCG 140-031 \nl
1727+2549 &    17 29 33.6 &   25 46 48 &   18.0 &   12$\times$12 &   121&   0.021 & I & m&   \nl
1729+2548 &    17 31 14.8 &   25 46 20 &   18.5 &   13$\times$12 &   25 &   0.020 & I & w&   \nl
1732+2414 &    17 34 49.5 &   24 12 29 &   18.5 &   10$\times$8 &   136 &   0.019 & C & w&   \nl
1732+2509 &    17 34 49.7 &   25 07 44 &   17.5 &   21$\times$12 &   121 &   0.022 & IP & m&   \nl
\tableline
\tablebreak
1733+2441 &    17 35 38.6 &   24 39 39 &   18.0 &   16$\times$8 &   87 &   0.018 & C & w&   \nl
1733+2554 &    17 35 14.3 &   25 52 31 &   17.5 &   11$\times$9 &   11 &   0.028 & C & s&   \nl
1734+2219 &    17 36 29.6 &   22 17 15 &   16.89$*$ &   21$\times$16 &   31 &   0.015 & S & m&   NPM1G +22.0588 \nl
1734+2322 &    17 36 36.3 &   23 21 08 &   17.61$*$ &   20$\times$14 &    94 &   0.025 & S & w&   NPM1G +23.0458 \nl
1735+2617 &    17 37 08.0 &   26 16 01 &   19.0 &   8$\times$6 &   73 &   0.014 & C & m&   \nl
1735+2622 &    17 37 48.0 &   26 21 18 &   16.51$*$ &   21$\times$17 &   46 &   0.022 & S & w&   NPM1G +26.0460 \nl
1736+2458 &    17 38 27.2 &   24 57 13 &   15.1$*$ &   54$\times$22 &   155 &   0.0209$*$ & S & w&   UGC 10926 \nl
1738+2544 &    17 40 14.4 &   25 43 05 &   18.0 &   11$\times$10 &   134 &   0.018 & I & m&   \nl
1739+2637 &    17 41 41.2 &   26 36 18 &   17.5 &   14$\times$12 &   35 &   0.030 & S & m&   \nl
1739+2639 &    17 41 46.8 &   26 38 00 &   17.5 &   26$\times$10 &   149 &   0.023 & S & m&   \nl
1740+2210 &    17 42 40.7 &   22 09 13 &   16.5 &   18$\times$13 &   124 &   0.043 & S & w&   \nl
1740+2351 &    17 42 45.1 &   23 50 30 &   17.5 &   22$\times$11 &   138 &   0.028 & S & w&   \nl
1742+2343 &    17 45 00.2 &   23 42 22 &   19.0 &   10$\times$9 &   133 &   0.014 & C & s&   \nl
1742+2634 &    17 44 40.1 &   26 33 25 &   19.5 &   10$\times$8 &    94 &   0.025 & C & w&   \nl
1744+2629 &    17 46 26.4 &   26 28 53 &   19.0 &   9$\times$8 &   169 &   0.024 & C & m&   \nl
1745+2235 &    17 47 18.5 &   22 34 41 &   17.0 &   17$\times$12 &   48 &   0.025 & S & w&   \nl
1746+2412 &    17 48 47.1 &   24 11 18 &   17.0 &   28$\times$15 &   40 &   0.028 & I & s&   \nl
\enddata
\tablenotetext{(2),(3)}{J2000.0 Equatorial Coordinates.}
\tablenotetext{(4)}{Data obtained from literature are marked with an asterisk}
\tablenotetext{(5)}{Sizes expressed in arcsec}
\tablenotetext{(7)}{Data obtained from literature are marked with an asterisk }
\tablenotetext{(8)}{C: Compact; I: Irregular; O: Oval; IP: Interacting
Pair; *: Stellar-like; S: Spiral [S$_f$: near face-on; S$_e$ near edge-on; S$_B$: Barred spiral]}
\tablenotetext{(9)}{Visual estimation of the H$\alpha$ emission
(w: weak; m: medium; s: strong)}
\normalsize
\end{deluxetable}


\end{document}